\documentclass{ws-procs961x669}            % default, citations in superscript
\begin{document}
\title{Gravitational anomalies, axions and a string-inspired running vacuum model in Cosmology}

\author{Nick E. Mavromatos$^{a,b}$}

\address{$^a$Theoretical Particle Physics and Cosmology Group, Physics Department, King's College London, \\
Strand, London WC2R 2LS, UK,
\vspace{0.1cm}\\
 $^{b}$ Physics Department, School of Applied Mathematical and Physical Sciences, National Technical University of Athens, \\
 9, Heroon Polytechneiou Str., Zografou Campus, Athens 157 80, Greece.\\
$^*$E-mail: nikolaos.mavromatos@cern.ch}

\begin{abstract}
I review a string-inspired cosmological model with gravitational anomalies in its early epochs, which is based on fields from the (bosonic) massless gravitational multiplet of strings, in particular gravitons and Kalb Ramond (KR), string-model independent, axions (the dilaton is assumed constant). I show how condensation of primordial gravitational waves, which are generared at the very early eras immediately after the big bang, can lead to inflation of the so called running vacuum model (RVM) type, without external inflatons. The role of the slow-roll field is played by the KR axion, but it does not drive inflation. The non-linearities in the anomaly terms do. Chiral fermionic matter excitations appear at the end of this RVM inflation, as a result of the decay of the RVM vacuum, and are held responsible for the cancellation of the primordial gravitational anomalies. Chiral anomalies, however, survive in the post-inflationary epochs, and can lead to the generation of a non perturbative mass for the KR axion, which could thus play the role of dark matter in this Universe. As a result of the condensed gravitational anomaly, there is a Lorentz-invariance violating KR axion background, which remains undiluted during the RVM inflation,
and can lead to baryogenesis through leptogenesis in the radiation era, in models with sterile right-handed neutrinos.
I also discuss the phenomenology of the model in the modern era, paying particular attention to linking it with a version of RVM, called type II RVM, which arguably can alleviate observed tensions in the current-epoch cosmological data.

\end{abstract}


\bodymatter

\section{Introduction}\label{sec:intro}

Although the plethora of the available cosmological data today~\cite{Planck} agree very well with the 
standard concordance model of Cosmology~\cite{pdg}, also known as $\Lambda$CDM, nonetheless there are compelling reasons for seeking alternative models, that go beyond it. Theoretically, given that the $\Lambda$CDM model is a global (de Sitter) solution of Einstein's general relativity (GR) theory, with a positive cosmological constant $\Lambda > 0$ added without explanation, 
one needs to seek microscopic frameworks of quantum gravity, which the GR can be embedded to, that could provide a detailed origin of the observed dark energy component in the current-Universe energy budget.\footnote{See, however, the arguments of Ref.~\refcite{subir} claiming that the ``observed'' dark energy might be an artefact of supernova data interpretation based on the assumption of homogeneity and isotropy of the universe at large (cosmological) scales, 
 in the context of the Friedman-Lemaitre-Robertson-Walker (FLRW) framework. In our approach here, the large-scale isotropy and homogeneity of the Universe is assumed.}
Moreover, one of the most 
important drawbacks of GR {\it per se} is that it is a non-renormalizable theory, and as such, it cannot provide by itself a framework for quantising gravity along with the rest of the fundamental  interactions in nature (assuming of course that gravity can  admit conventional quantization; for alternative views, treating gravity as an entropic force, associated with changes of information pertaining to the positions of material bodies, see Ref.~\refcite{verlinde}). 
On the observational side, there appear to exist discrepancies (``tensions''~\cite{tensions,sol2}) between the results of data analyses from Planck Collaboration, based on $\Lambda$CDM~\cite{Planck}, and local direct measurements of the Hubble parameter today, the so called $H_0$ tension~\cite{Htension,Htension1,Htension2}, but also tensions in the 
growth-of-galaxies data, the so-called $\sigma_8$ tension~\cite{s8tension}. Thus, although such tensions could admit more mundane astrophysical explanations~\cite{efst,dust,dust2,sol3}, or even disappear in future data, being due to statistical uncertainties, nonetheless, several theoretical attempts are currently on the way in order to provide an explanation for them or alleviate them. 

One phenomenological framework, which deals successfully with such an alleviation of both the $H_0$ and the $\sigma_8$ tensions, is a 
mildly modified version of the running vacuum model of cosmology (RVM)~\cite{rvm0,rvm,rvmfoss,rvmevol,rvmevol2,rvmsola}, the so-called type II RVM~\cite{solaepl}. 
In this talk, I will review a microscopic framework for obtaining the RVM, actually a version of RVM somewhat analogous to the aforementioned type II RVM, which pertains to a string-inspired~\cite{string} cosmological model with {\it gravitational anomalies} in the early Universe~\cite{bas0,basilakos2,basilakos,ms2,ms1,ms0}.

The structure of the talk is as follows: in the next section \ref{sec:sRVM}, I describe briefly the RVM framework, and then the string-inspired gravitational theory, with gravitational anomalies coupled to {\it massless} string-model independent (Kalb-Ramond (KR)) axions fields, which can give rise to an RVM inflation, without external inflatons. 
Only bosonic fields from the massless gravitational string multiplet enter the early-Universe era string-effective action as external fields. 
Primordial gravitational waves can lead to condensation of the gravitational anomalies, which in turn induces the RVM inflation.
In section \ref{sec:postinfl}, I describe briefly the post inflationary epochs, placing the emphasis on how a  gravitational-anomaly-induced  Lorentz-violating (LV) KR axion background, which remans undiluted at the end of the RVM inflationary era, 
can lead to leptogenesis in models with right-handed sterile neutrinos (RHN) in their spectra. I also explain how the stringy KR axion field can play the r\^ole of Dark matter, by acquiring a non-perturbative mass during the Quantum ChromoDynamics (QCD) epoch of the Universe. In section \ref{sec:modern}, I speculate on potential links of this stringy RVM effective theory to a modified version of RVM, somewhat analogous (but with crucial differences) to the aforementioned type-II RVM. The crucial modifications are due to quantum-graviton-mode integration in the respective path integral of the string-inspired theory, and could lead to observable departures from $\Lambda$CDM and potential alleviation of tensions  in the current-era data. Finally, section \ref{sec:concl} contains our conclusions and outlook. 

\section{Running Vacuum Model (RVM) Framework, its String-theory embedding  and Inflation from Gravitational Anomalies \label{sec:sRVM}}

I commence the discussion by first summarising briefly the basic features of the RVM framework. 
This is a phenomenologically successful effective cosmological framework, which leads to observable, 
in principle, deviations from $\Lambda$CDM~\cite{rvmevol2,rvmsola}, and alleviation of the current cosmological tensions~\cite{solaepl}. It describes a smooth 
evolution of the Universe from a dynamical inflationary phase till the present era~\cite{rvmevol}.  

\subsection{Features of RVM cosmology \label{sec:rvmcosmol}}

The basic feature of the RVM cosmology is that the vacuum energy density is a function of even powers of the Hubble parameter at a given era of the Universe, $H(t)$ (with $t$ the cosmic time), due to general covariance. To describe the entire evolution of the Universe from inflation till the present epoch, it is sufficient to truncate this series to quartic powers of $H(t)$, thus writing:\cite{rvm0,rvm,rvmfoss,rvmevol}:
\begin{equation}\label{rLRVM}
\rho^{\Lambda}_{\rm RVM}(H) = \frac{\Lambda(H)}{\kappa^2}=
\frac{3}{\kappa^2}\left(c_0 + \nu H^{2} + \alpha
\frac{H^{4}}{H_{I}^{2}} + \dots \right) \, >\, 0\;,
\end{equation}
with $\kappa$ the four-dimensional gravitational
constant $\kappa = \sqrt{8\pi\, {\rm G}} = M_{\rm Pl}^{-1}$, with $M_{\rm Pl} =  2.43 \times 10^{18}$~GeV  (we work in natural units $\hbar=c=1$ throughout this work), $H_I \sim 10^{-5} M_{\rm Pl}$, 
the inflationary scale, as inferred from the latest Planck-Collaboration data~\cite{Planck},  
and $c_0 \ge 0 $,  $\nu$ and $\alpha$ constants, throughout the 
Universe evolution. In the conventional RVM, $\nu > 0$ and $\alpha >0$, while $c_0 > 0$ is the current-epoch cosmological constant~\cite{pdg,Planck}. The $\dots$ in \eqref{rLRVM} denote terms of higher orders in $H^2(t)$. In general there is also dependence on $\dot H$ (with the overdot denoting cosmic-time-$t$ derivative), but for a given era of the Universe, one can express
$\dot H$ in terms of the cosmic deceleration of that era (assumed approximately constant) and of $H^2$, hence the expression \eqref{rLRVM} suffices for our purposes here.  The defining feature of the RVM is its de-Sitter type equation of state~\cite{rvm0}, 
\begin{align}\label{rvmeos}
p^{\rm vac} _{\rm RVM} (H(t)) = - \rho^{\rm vac}_{\rm RVM} (H(t)) \, <\, 0,
\end{align}
despite the fact that the pressure $p_{\rm RVM}^{\rm vac}$ and energy $\rho_{\rm RVM}^{\rm vac}$ densities are time-dependent functions of $H(t)$. We note for completeness that a similar feature characterises the super-critical (Liouville) string cosmologies, where the cosmic time is identified with (the zero mode of) a world-sheet renormalization-group scale~\cite{emn}. 

From the conservation of the total stress tensor of the vacuum plus radiation and matter excitations, the latter being characterised by an equation of state $w_m$, one can easily arrive at the following evolution  equation for the Hubble parameter~\cite{rvmevol}:
\begin{equation}\label{evol}
\dot H + \frac{3}{2} \, (1 + \omega_m) \, H^2 \, \Big( 1 - \nu - \frac{c_0}{H^2} - \alpha \, \frac{H^2}{H_I^2} \Big) =0~.
\end{equation}
At early stages of the Universe, one may ignore the $c_0$ term, thus arriving at a 
 solution for $H(a)$ as a function of the scale factor $a$ (in units of the present-era scale factor):\cite{rvmevol,rvmevol2}
\begin{equation}\label{HS1}
 H(a)=\left(\frac{1-\nu}{\alpha}\right)^{1/2}\,\frac{H_{I}}{\sqrt{D\,a^{3(1-\nu)(1+\omega_m)}+1}}\,,
\end{equation}
where $D>0$ is an integration constant.  For the early Universe,  one may assume without loss of generality that
$D\,a^{3(1-\nu)(1+\omega_m)} \ll 1$, which from \eqref{HS1} leads to
an (unstable) dynamical early de Sitter phase, characterised by
an approximately constant $H_{\rm de~Sitter}  \simeq \left(\frac{1-\nu}{\alpha}\right)^{1/2}\,\, H_{I}$. This inflationary era is due to the non-linear terms $H^4$ in \eqref{rLRVM}, and is not characterised by external inflaton fields. 

After inflation, the RVM evolution \eqref{evol} is characterised by radiation-dominance ($w_m=1/3)$ and, subsequently, by a matter-dominated
($w_m=0$) era, both occurring for considerably larger values of the scale factor than the RVM inflation. In such cases,  the corresponding quantities
$D\,a^{3(1-\nu)(1+\omega_m)} \gg 1$. One can therefore determine the RVM vacuum energy density \eqref{rLRVM} during radiation as~\cite{rvmevol,sugrarvm}  
$\rho_{\rm RVM}^{\rm vac~radiation~era} \simeq \frac{3H_I^2}{\kappa^2\, \alpha} \frac{1}{(1 + D a^4)^2}$,
and the radiation energy density $\rho_{\rm rad} (a) \simeq \frac{3H_I^2}{\kappa^2\, \alpha} \frac{D a^4}{(1 + D a^4)^2}$,  
thus recovering the familiar $a^{-4}$ scaling for $D a^4 \gg 1$. As the evolution continues, matter starts to dominate ($\omega_m=0$), and at that point the cosmological constant term $c_0$  can no longer be ingored, leading eventually to a modern-era 
scaling of the Hubble parameter $H \ll H_I$ (obtained as the solution of the evolution equation \eqref{evol} at this era)~\cite{rvmevol,sugrarvm}: $H^2_{\rm modern} (a) \simeq \frac{H_0^2}{1-\nu} \Big[ ( 1 - \Omega_\Lambda^0) \, a^{-3(1-\nu)} + \Omega_\Lambda^0 - \nu \Big]$, with $H_0$ the present-day Hubble parameter, and $\Omega_\Lambda^0 = 
\frac{c_0 + H_0^2 \nu}{H_0^2}$. Notice that the presence of the parameter $\nu > 0$ implies observable deviations from the
$\Lambda$CDM concordance model of cosmology~\cite{rvmevol2,rvmsola} (which is recovered in the limit $\nu=0$, in which case $\Omega_\Lambda^0$ coincides with the present-era cosmological constant (in units of the current-era critical density)).

We remark at this stage that the non-linearities-induced inflation  \eqref{HS1} is different~\cite{rvmevol2,sugrarvm,ms2} from Starobinsky's model of inflation~\cite{staro}, which is also due to conformal-(``trace'')-anomaly-induced non-linear curvature corrections to Einstein's GR theory. Starobinsky's model does not have in its energy density the crucial $H^4$ term of the RVM ({\it cf.} \eqref{rLRVM}). As discussed in Refs.~\refcite{rvmevol2,ms2}, it is an inflation characterised mainly by  $\dot H \simeq$ constant rather than $H \simeq $ constant, which is the case of RVM. 
The crucial for inflation $H^4$ term of the RVM energy density \eqref{rLRVM} is also missing in the case of a quantum field theory of a scalar field non-minimally coupled to a gravity background~\cite{rvmqft}.
As we shall discuss below, a direct (geometric in origin) induction of an RVM $H^4$ term in the corresponding energy density 
can be achieved through {\it condensates} of {\it gravitational anomalies}, induced by primordial gravity waves~\cite{stephon}, in the string-inspired model for cosmology discussed in Refs.~\refcite{bas0,basilakos2,basilakos,ms2,ms1,ms0}, which 
we now come to describe briefly.

\subsection{String-inspired RVM, Gravitational Anomalies and Inflation \label{sec:stringyrvm}}

The basic assumption~\cite{basilakos2} towards the construction of the string-inspired cosmology model that will lead to RVM, is that at early stages of the Universe, only fields from the bosonic massless gravitational multiplet of strings~\cite{string} appear as external fields. Assuming constant dilatons, this assumption implies that the effective action in the four dimensional space time, after string compactification, will consist only of gravitons and antisymmetric tensor fields $B_{\mu\nu}=-B_{\nu\mu}$, $\mu,\nu=0, \dots 3$. As a result of an appropriate U(1) gauge symmetry in the closed string sector~\cite{string}: $B_{\mu\nu} \, \to \, B_{\mu\nu} + \partial_\mu \theta_\nu (X) - \partial_\nu \theta_\mu (X)$, where $\theta_\mu(X)$ are gauge parameters, the field $B_{\mu\nu}$ appears in the effective four space-time dimensional action only through its field strength. Due to the Green-Schwarz anomaly cancellation mechanism~\cite{string}, the latter is modified by appropriate gauge and gravitational Chern-Simons terms. In our model, only the latter are present in the early Universe~\cite{basilakos2}, and as such the modified field strength 
$\mathcal H_{\mu\nu\rho}$ of the field $B_{\mu\nu}$ reads (in differential form language)~\cite{string}: $\mathbf{{\mathcal H}} = \mathbf{d} \mathbf{B} + \frac{\alpha^\prime}{8\, \kappa} \, \Omega_{\rm 3L}$, with  $\alpha^\prime = M_s^{-2}$ the Regge slope, $M_s$ the string mass scale, and the gravitational Chern Simons term is defined as 
$\Omega_{\rm 3L} = \omega^a_{\,\,c} \wedge \mathbf{d}\, \omega^c_{\,\,a} + \frac{2}{3}  \omega^a_{\,\,c} \wedge  \omega^c_{\,\,d} \wedge \omega^d_{\,\,a}$, 
where $\wedge$ denotes the exterior product among forms, $\mathbf{d}$ is the exterior derivative one form. 
and $\omega_{\mu\,\, b}^a$ is the spin connection, with Latin indices $a,b=0, \dots 3$ denoting tangent space indices. The field strength satisfies a Bianchi identity:
\begin{align}\label{modbianchi}
 \varepsilon_{abc}^{\;\;\;\;\;\;\mu}\, {\mathcal H}^{abc}_{\;\;\;\;\;\; ;\mu}
 =  \frac{\alpha^\prime}{32\, \kappa} \, \sqrt{-g}\, R_{\mu\nu\rho\sigma}\, \widetilde R^{\mu\nu\rho\sigma} 
 \end{align}
where the semicolon denotes covariant derivative with respect to the standard
Christoffel connection, 
$ \varepsilon_{\mu\nu\rho\sigma} $ is the gravitationally covariant Levi-Civita tensor density, totally antisymmetric in its indices,  
and $\widetilde R_{\mu\nu\rho\sigma} = \frac{1}{2} \varepsilon_{\mu\nu\lambda\pi} R_{\,\,\,\,\,\,\,\rho\sigma}^{\lambda\pi}$ is the dual Riemann curvature tensor (we follow the conventions of Ref.~\refcite{basilakos2}). 
To lowest (quadratic) order in derivatives, the target-space effective action (for constant dilatons) reads:
\begin{align}\label{sea}
S_B  =\; \int d^{4}x\sqrt{-g}\Big( \dfrac{1}{2\kappa^{2}} [-R  - \frac{1}{6}\,  {\mathcal H}_{\lambda\mu\nu}{\mathcal H}^{\lambda\mu\nu} + \dots \Big),
\end{align}
 with the $\dots$ representing higher-derivative terms. We mention at this stage the r\^ole of $\mathcal H_{\mu\nu\rho}$ as 
 a totally antisymmetric component of torsion in string-inspired effective gravitational theories~\cite{string,kaloper}, in the sense 
 that the action \eqref{sea} can be expressed in terms of a generalised curvature scalar with respect to such a torsion.
 As we shall describe below, the torsion is associated with an axion field~\cite{kaloper}. 
  
Indeed, on implementing \eqref{modbianchi}  as a constraint in the respective path-integral of the action \eqref{sea4} 
by means of a pseudoscalar Lagrange mutl[plier field $b(x)$, and integrating out $\mathcal H$, one obtains the 
effective action~\cite{kaloper,svrcek}
\begin{align}\label{sea4}
S^{\rm eff}_B &= \; \int d^{4}x\sqrt{-g}\Big[ -\dfrac{1}{2\kappa^{2}}\, R + \frac{1}{2}\, \partial_\mu b \, \partial^\mu b +   \sqrt{\frac{2}{3}}\,
\frac{\alpha^\prime}{96 \, \kappa} \, b(x) \, R_{\mu\nu\rho\sigma}\, \widetilde R^{\mu\nu\rho\sigma} + \dots \Big] \nonumber \\
&= \; \int d^{4}x\, \sqrt{-g}\Big[ -\dfrac{1}{2\kappa^{2}}\, R + \frac{1}{2}\, \partial_\mu b \, \partial^\mu b  -
 \sqrt{\frac{2}{3}}\,
\frac{\alpha^\prime}{96 \, \kappa} \, {\mathcal K}^\mu (\omega)\, \partial_\mu b(x)   + \dots \Big],
\end{align}
where we took into account that the gravitational anomaly is a total derivative
$\sqrt{-g} \, R_{\mu\nu\rho\sigma}\, \widetilde R^{\mu\nu\rho\sigma} = \sqrt{-g} \, {\mathcal K}^\mu (\omega)_{;\mu} = \partial_\mu \Big(\sqrt{-g} \, {\mathcal K}^\mu (\omega) \Big)
= - 2 \, \partial_\mu \Big[\sqrt{-g}\, \varepsilon^{\mu\nu\alpha\beta}\, \omega_\nu^{ab}\, \Big(\partial_\alpha \, \omega_{\beta ab} + \frac{2}{3}\, \omega_{\alpha a}^{\,\,\,\,\,\,\,c}\, \omega_{\beta cb}\Big)\Big]$. One observes from \eqref{sea4} that the Lagrange multiplier field $b(x)$ became a fully dynamical axion-like field, the so called Kalb-Ramond (KR) or string-model-independent axion~\cite{kaloper,svrcek}.  Classically, as follows from the saddle-point of the path integral over $\mathcal H$, one has the relation~\cite{kaloper}: $3\sqrt{2} \, \partial_\sigma \bar b = -\sqrt{-g} \, \epsilon_{\mu\nu\rho\sigma} \, \overline{\mathcal H}^{\mu\nu\rho}$, where the bar denotes classical fields. Hence the association of the torsion with the axion field $b(x)$.

The effective action \eqref{sea4} forms the basis of our cosmological model~\cite{basilakos2}. In a FLRW 
space-time background the gravitational anomaly terms vanish~\cite{jackiw}. This is not true, however,  
in the presence of primordial 
gravity waves (GW) perturbations, which violate CP symmetry: 
\begin{align}\label{metric2}
 ds^2 = dt^2 - a^2(t) \Big[(1 - h_+(t,z))\, dx^2 + (1 + h_+(t,z))\, dy^2 + 2h_\times (t,z)\, dx\, dy + dz^2 \Big],
 \end{align}
using standard notation for the graviton polarizations of the GW.
In the metric \eqref{metric2}, the gravitational anomaly term is {\it non zero}~\cite{stephon}. In fact, on assuming an inflationary space time with constant Hubble parameter $H \simeq$ constant, one can compute the anomaly condensate in the presence of GW CP-violating perturbations~\cite{stephon}:
 \begin{align}\label{rrt2}
  \langle R_{\mu\nu\rho\sigma}\, \widetilde R^{\mu\nu\rho\sigma} \rangle  &\simeq  
  \frac{16}{a^4} \, \kappa^2\int^\mu \frac{d^3k}{(2\pi)^3} \, \frac{H^2}{2\, k^3} \, k^4 \, \Theta  = \frac{1}{\pi^2} \Big(\frac{H}{M_{\rm Pl}}\Big)^2 \, \mu^4\, \Theta   \nonumber \\
 &= \frac{2}{3\pi^2} \frac{1}{96 \times 12} \,  \Big(\frac{H}{M_{\rm Pl}}\Big)^3 \, \Big(\frac{\mu}{M_{\rm Pl}}\Big)^4 \,  M_{\rm Pl}\, \times \, \, {\mathcal K}^0 (t)\,,
\end{align}
to leading order in the slow-roll parameter 
\begin{align}\label{theta}
 \Theta = \sqrt{\frac{2}{3}}\, \frac{\alpha^\prime \, \kappa}{12} \, H \,  {\dot {\overline b}} \, \ll \, 1~,
  \end{align}
with the overdot denoting derivative with respect to the cosmic time $t$. The quantity $\mu$ is an UltraViolet (UV) cutoff 
in the momenta of the graviton modes. 

Assuming isotropy and homogeneity in the Universe, the condensate \eqref{rrt2} implies the following solution of the Euler-Lagrange equation of  the KR axion $b(t)$~\cite{basilakos2}: 
\begin{align}\label{krbeom2}
\dot{\overline{b}}  =  \sqrt{\frac{2}{3}}\, \frac{\alpha^\prime}{96 \, \kappa} \, {\mathcal K}^{0} \simeq {\rm constant},
\end{align} 
provided  $\mu/M_s \simeq 15 \, \Big(M_{\rm Pl}/H\Big)^{1/2}$. Combining the slow-roll condition \eqref{theta}, with the transplanckian conjecture, {\it i.e.} the absence of transplanckian graviton modes ($\mu \le M_{\rm Pl}$), one can show that 
for the inflationary Hubble parameter $H = H_I = 10^{-5} \, M_{\rm Pl}$, as dictated by data~\cite{Planck}, the string mass scale is restricted to the range~\cite{ms1,ms0}:
\begin{align}\label{stringscale}
2.6 \times 10^{-5}\, M_{\rm Pl}  \lesssim M_s \lesssim 10^{-4}\, M_{\rm Pl}.
\end{align}

Eq.~\eqref{krbeom2} admits a spontaneous-Lorentz-violating (LV) KR axion background solution, parametrised as~\cite{basilakos2,ms2}:
\begin{align}\label{slowrollbint}  
b(t) = \overline b(0) + \sqrt{2\epsilon} H  M_{\rm Pl} \, t, \quad H \simeq {\rm constant}, 
\end{align}
with $\overline b(0)$ a boundary condition for the field $b(x)$ at the onset of inflation at $t=0$. For phenomenological
reasons~\cite{Planck} we may take~\cite{basilakos2} $\epsilon = \mathcal O(10^{-2})$. Thus, the KR axion is a slow-roll field, but, as we shall discuss next, it does not drive inflation.

To this end, we remark that the condensate \eqref{rrt2} implies a condensate of the quantity~\cite{basilakos2}:  $\langle b(t)  R_{\mu\nu\rho\sigma} \, \widetilde R^{\mu\nu\rho\sigma} \rangle$, which remains approximately constant during the entire duration of inflation, thus corresponding to a de-Sitter-type (cosmological-constant) contribution to the effective target-space action \eqref{sea4}, provided that 
\begin{align}\label{b0}
\frac{|\overline b(0)|}{M_{\rm Pl}}  \gg \, \sqrt{2\, \epsilon} \,  \mathcal N = \mathcal O(10), \quad \overline b(0) < 0,
\end{align}
where the end of inflation~\cite{inflation} has been set to $t_{\rm inf} H \simeq \mathcal N$, with 
$\mathcal N$ the number of e-foldings of the inflationary era, which phenomenologically~\cite{Planck} lies in the range 
$\mathcal N=60-70$. 

Expanding the effective action \eqref{sea4} about such de-Sitter background configurations, one can show that the equation of state of the total energy $\rho_{\rm total} =  \rho^b + \rho^{\rm gCS} + \rho^{\rm condensate}$  
and pressure $p_{\rm total} = p^b + p^{\rm gCS} + p^{\rm condensate}$
densities of this fluid, consisting of contributions from the KR axion (superscript $b$), 
fluctuations of the gravitational anomaly (Chern-Simons) term (superscript gCS) and the condensate term (superscript ``condensate''), assumes an RVM form~\cite{ms1}:
\begin{align}\label{rvmeos}
p_{\rm total}  = - \rho_{\rm total} \, <\, 0,
\end{align}
with the total energy density being dominated by the $H^4$ term during the inflationary era 
\begin{align}\label{toten}
0 <  \rho_{\rm total}  \simeq  3\kappa^{-4} \, \Big[ -1.65 \times 10^{-3} \Big(\kappa\, H \Big)^2
+ \frac{\sqrt{2}}{3} \, |\overline b(0)| \, \kappa \, \times {5.86\, \times} \, 10^6 \, \left(\kappa\, H \right)^4 \Big]
\end{align}
under the condition \eqref{b0}. The reader should notice that the coefficient of the $H^2$ term is negative, due to the contribution of the anomalous gravitational Chern-Simons terms, and this is a difference from the conventional RVM form \eqref{rLRVM}.
We stress that it is the dominance of the condensate term that makes the total energy density \eqref{toten} positive, and thus leads to a proper RVM fluid. Without the formation of this condensate, the equation of state of the remaining terms curiously acquire the form of an RVM-like (i.e. de Sitter-type) ``phantom-matter''~\cite{ms1,phantom,phantom2}:
\begin{align}\label{bgCS}
p^b + p^{\rm gCS} =  - (\rho^b + \rho^{\rm gCS})>0~,
\end{align}
violating the weak energy conditions.\footnote{The computation leading to \eqref{bgCS} is based on the fact that the graviton variations of the anomalous gravitational Chern-Simons terms in \eqref{sea4} yield the Cotton tensor 
$C_{\mu\nu}$~\cite{jackiw}: 
$$\delta \Big[ \int d^4x \sqrt{-g} \, b \, R_{\mu\nu\rho\sigma}\, \widetilde R^{\mu\nu\rho\sigma} \Big] = 4 \int d^4x \sqrt{-g} \, {\mathcal C}^{\mu\nu}\, \delta g_{\mu\nu} = -4 \int d^4x \sqrt{-g} \, {\mathcal C}_{\mu\nu}\, \delta g^{\mu\nu},$$
where 
$${\mathcal C}^{\mu\nu} \equiv  -\frac{1}{2}\, \Big[v_\sigma \, \Big( \varepsilon^{\sigma\mu\alpha\beta} R^\nu_{\, \, \beta;\alpha} + \varepsilon^{\sigma\nu\alpha\beta} R^\mu_{\, \, \beta;\alpha}\Big) + v_{\sigma\tau} \, \Big(\widetilde R^{\tau\mu\sigma\nu} +
\widetilde R^{\tau\nu\sigma\mu} \Big)\Big],$$ with $v_{\sigma} \equiv \partial_\sigma b = b_{;\sigma}, \,\,v_{\sigma\tau} \equiv  v_{\tau; \sigma} = b_{;\tau;\sigma}.$
This implies that the Einstein's equations stemming from \eqref{sea4} read: 
$R^{\mu\nu} - \frac{1}{2}\, g^{\mu\nu} \, R  - \,  {\mathcal C}^{\mu\nu} = \kappa^2 \, T^{\mu\nu}_b$, where $T^{\mu\nu}_b$ is the KR axion stress tensor. Taking into account conservation properties of the Cotton tensor\cite{jackiw}, 
${\mathcal C}^{\mu\nu}_{\,\,\,\,\,\,\,;\mu} = \frac{1}{8}\, v^\nu \, R^{\alpha\beta\gamma\delta} \, \widetilde R_{\alpha\beta\gamma\delta}$, we observe that there is an exchange of energy
between the KR axion and the gravitational anomaly, when $\mathcal C_{\mu\nu}$ is non trivial, as is the case 
of GW perturbations of the metric ({\it cf.} \eqref{metric2}). We note that classically the Cotton tensor is traceless $g^{\mu\nu} \mathcal C_{\mu\nu}=0$. However, the formation of the condensate $\langle b(t)  R_{\mu\nu\rho\sigma} \, \widetilde R^{\mu\nu\rho\sigma} \rangle$, as a result of \eqref{rrt2}, introduces in a sense a quantum-gravity-induced 
trace of $\langle \mathcal C^\mu_{\,\mu}\rangle$, which leads to the addition of a de Sitter type cosmological-constant term in the effective action \eqref{sea4}~\cite{basilakos2}.} 
Such exotic matter can be used for the stabilisation of traversable wormholes~\cite{worm}, so in this respect our anomalous gravitational theory, in conditions in which the condensate $\langle b(x)\, R_{\mu\nu\rho\sigma} \, \widetilde R^{\mu\nu\rho\sigma} \rangle$ is not formed, might find applications to this problem as well.

The  dominance of the non-linear $H^4$ term in \eqref{toten} at early eras of this string-inspired Universe, will lead to an RVM type inflation \eqref{HS1}, as we discussed in the previous subsection \ref{sec:rvmcosmol}. It is in this sense that an inflationary phase arises dynamically in a self consistent way in our approach, so our assumption of a constant $H$ in the computation of \eqref{rrt2} is justified {\it a posteriori}.

\subsection{Potential origin of GW: the dynamically-broken Supergravity example \label{sec:sugra}} 

So far, we have assumed the existence of primordial GW perturbations \eqref{metric2}, without examining in detail their microscopic origin. In Refs.~\refcite{ms2,ms1}, we have discussed several scenarios which lead to the creation of GW in a pre-RVM-inflationary era of our string-inspired cosmology. One of the most relevant ones to our model, 
consistent with the basic assumption that only fields from the massless gravitational string multiplet appear as external fields at the very early stages of the Universe, is the scenario considering an underlying superstring theory, whose low energy limit 
will result in a supergravity model. Although the precise supergravity model depends crucially on the microscopic higher-dimensional string theory considered, for our qualitative purposes here, we may consider a simplified but highly non-trivial example for our (3+1)-dimensional gravitational theory at the very early stages of our cosmology, that of $N=1$ (3+1)-dimensional supergravity~\cite{N1sugra}. 
\begin{figure}[!h]
\begin{center}
\includegraphics[width=4.5in]{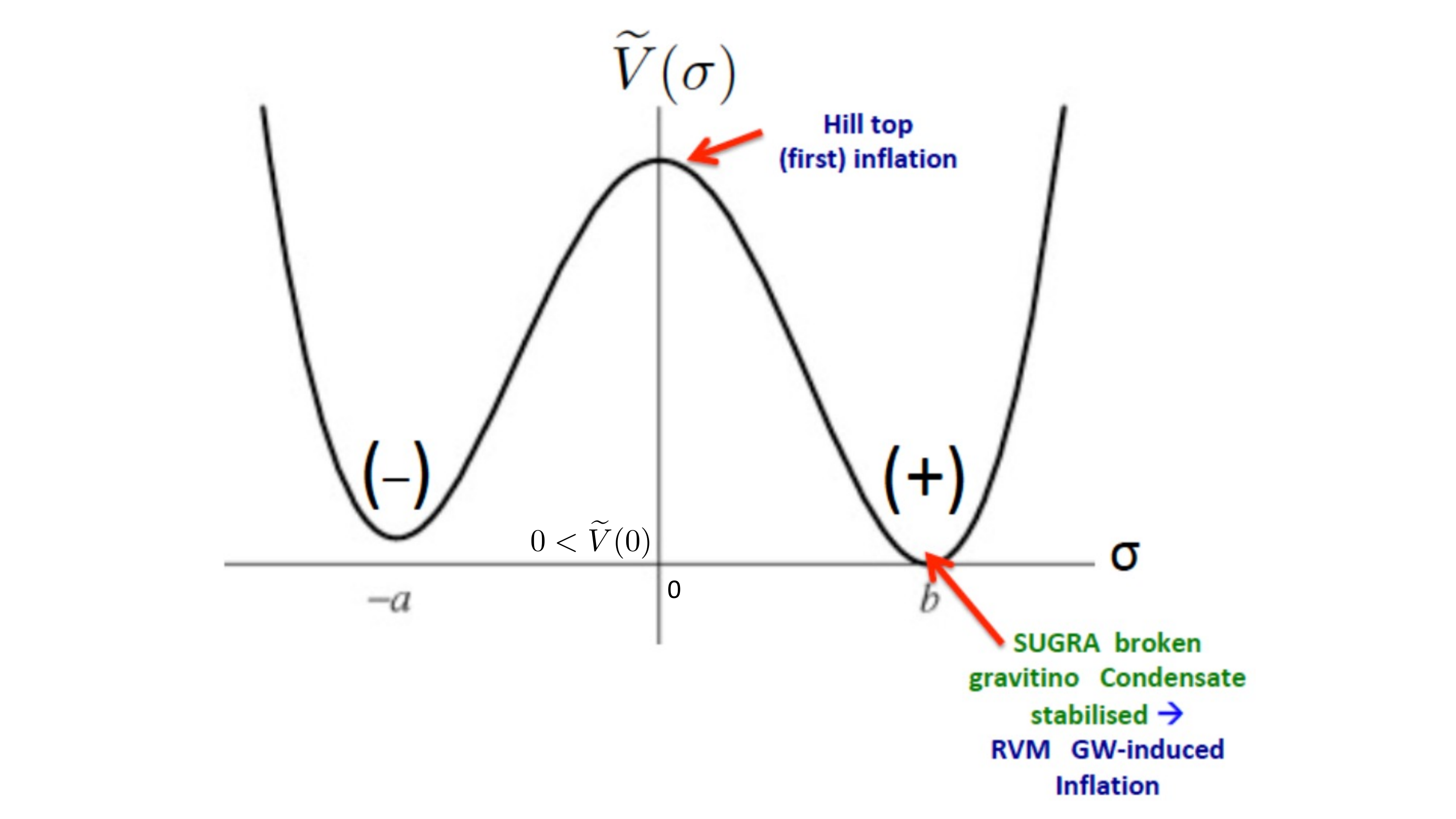}
\end{center} 
\caption{The biased double-well gravitino potential $\widetilde V(\sigma)$ for dynamical supergravity breaking that could characgterise the early epochs of the string-inspired RTVM Cosmology~\protect\cite{ms2}.
The bias between the two $(\pm)$ vacua is assumed to be due to percolation effects in the early Universe~\protect\cite{ovrut,ovrut2}, and leads to the formation of unstable domain walls, whose collapse or collisions lead to GW. There is also a first hill-top inflation near the origin $\sigma=0$.}
\vspace{-0.3cm}
\protect\label{fig:sugra}
\end{figure}

The model contains massless spin-2 gravitons, $g_{\mu\nu}(x)$, $\mu,\nu=0, \dots 3$,  and their spin-3/2 supersymmetric partners, the gravitinos, $\psi_\mu(x)$. The
supergravity can be broken dynamically in this model as a consequence of gravitino condensation~\cite{houston,houston2}. The corresponding effective potential $\widetilde V(\sigma)$ of the gravitino-condensate scalar field $\sigma = \langle \overline \psi_\mu \psi^\mu \rangle $, evaluated at one loop in a local de Sitter space-time background~\cite{tseytlin}, after graviton and gravitino Euclidean path-integration, assumes a double-well shape (see fig.~\ref{fig:sugra}), defining two local minima, at each of which the potential takes on non-negative values, compatible with the breaking of global and local supersymmetry. There could be a statistical bias, in the sense of 
unequal occupation numbers of these two local minima/``vacua'', as a result of percolation effects~\cite{ovrut,ovrut2}. 
Such a situation leads to the formation of ``biased'' domain walls in the theory, whose non-spherical collapse or collision, leads to GW. The $N=1$ supergravity model is also characterised by a very early hill-top inflationary phase~\cite{ellis}, which is 
not necessarily slow roll, and has no observable consequences. 

This first inflationary phase is consistent with the study of the theory in a local de Sitter (LDS) background, in the absence of a mass parameter~\cite{tseytlin}, given the existence of an appropriate general coordinate transformation~\cite{lanczos}  that maps the LDS space-time without a mass parameter to a cosmological (global) de Sitter one (GDS), and leaves the effective action invariant.\footnote{The local de Sitter-Schwarzschild metric with a mass parameter $M$ is described by the following invariant element
\begin{align}\label{SDS}
 ds^2= \left(1-2\frac{M}{\bar{r}} - \frac{\Lambda}{3}\, \bar{r}^2 \right) c^2 dT^2 - \left(1-2\frac{M}{\bar{r}} - \frac{\Lambda}{3}\, \bar{r}^2 \right)^{-1} d\bar{r}^2
          -\bar{r}^2 (d\bar{\theta}^2 +  \rm{sin}^2\bar{\theta} d\bar{\phi}^2)
\end{align}
in de Sitter-Schwarzschild coordinates. In the case $M=0$, as required by the isotropy and homogeneity of space, 
the following transformation~\cite{lanczos}
\begin{align}\label{lanczostr}
x^\mu &\equiv \{ c\,T, {\bar r}, {\bar \theta}, {\bar \phi} \} \, \rightarrow  \, x^{\prime \, \mu} \equiv \{c\,t, r, {\bar \theta}, {\bar \phi} \}  \, \,\,{\rm (comoving~frame)} \, : \nonumber \\
t &= T + \frac{1}{2 \, H} \, {\rm ln} \Big(1 - H^2 \, {\bar r}^2 \Big), \quad
r = \frac{{\bar r}}{\sqrt{1 - H^2 \, {\bar r}^2 }}\, e^{-H\, T} = {\bar r} \, e^{-H t},  \quad H^2 \equiv \frac{\Lambda}{3} \, > \, 0~.
\end{align}
where $t$ and $r$ denote co-moving frame time and radial space coordinates, respectively, 
maps the metric \eqref{SDS} to a standard cosmological de Sitter space-time:
\begin{align}\label{inflation}
ds^2 = c^2 \, dt^2 - a(t)^2 \big[ dr^2 + r^2 d\Omega^2 \big], \quad d\Omega^2 = d\bar{\theta}^2 +  \rm{sin}^2\bar{\theta} d\bar{\phi}^2~,
\end{align}
where $a(t)=e^{\sqrt{\frac{\Lambda}{3}}\, t} \equiv e^{H\, t} $ ($H=$ constant), is the exponentially expanding scale factor of the de-Sitter/FLRW (inflationary) Universe. The above result is exact, valid for every non-negative value of the cosmological constant $\Lambda \ge 0$. Notice that in this form of the metric there is no preferred origin of space, as appropriate for a space-time generated by a uniform homogeneous and isotropic fluid. 
The apparent existence of an origin at zero for the radial coordinate in the local form of the pure de-Sitter metric (\ref{SDS}) with $M=0$ is thus an artefact of an inappropriate choice of coordinates, much like the internal observer de Sitter horizon at $\sqrt{\Lambda/3}$. }
 
This hill-top inflation is a different phase from our RVM inflation. In fact, it ensures that any spatial inhomogeneities or isotropies are washed out well before entrance into the RVM inflationary phase, so the assumptions and calculations of Refs.~\protect\refcite{basilakos2,ms2} are formally justified. The RVM inflation occurs~\cite{ms2} at the phase where the gravitino has settled in its lowest minimum (see fig.~\ref{fig:sugra}), is massive and the supergravity is dynamically broken. This phase coincides with the exit phase from the hill-top inflation~\cite{ellis}. This exit phase is characterised by the creation of the biased domain walls and GW, as a consequence of wall collisions. The massive gravitino can be integrated out of the path integral.
\begin{figure}[!h]
\begin{center}
\includegraphics[width=5in]{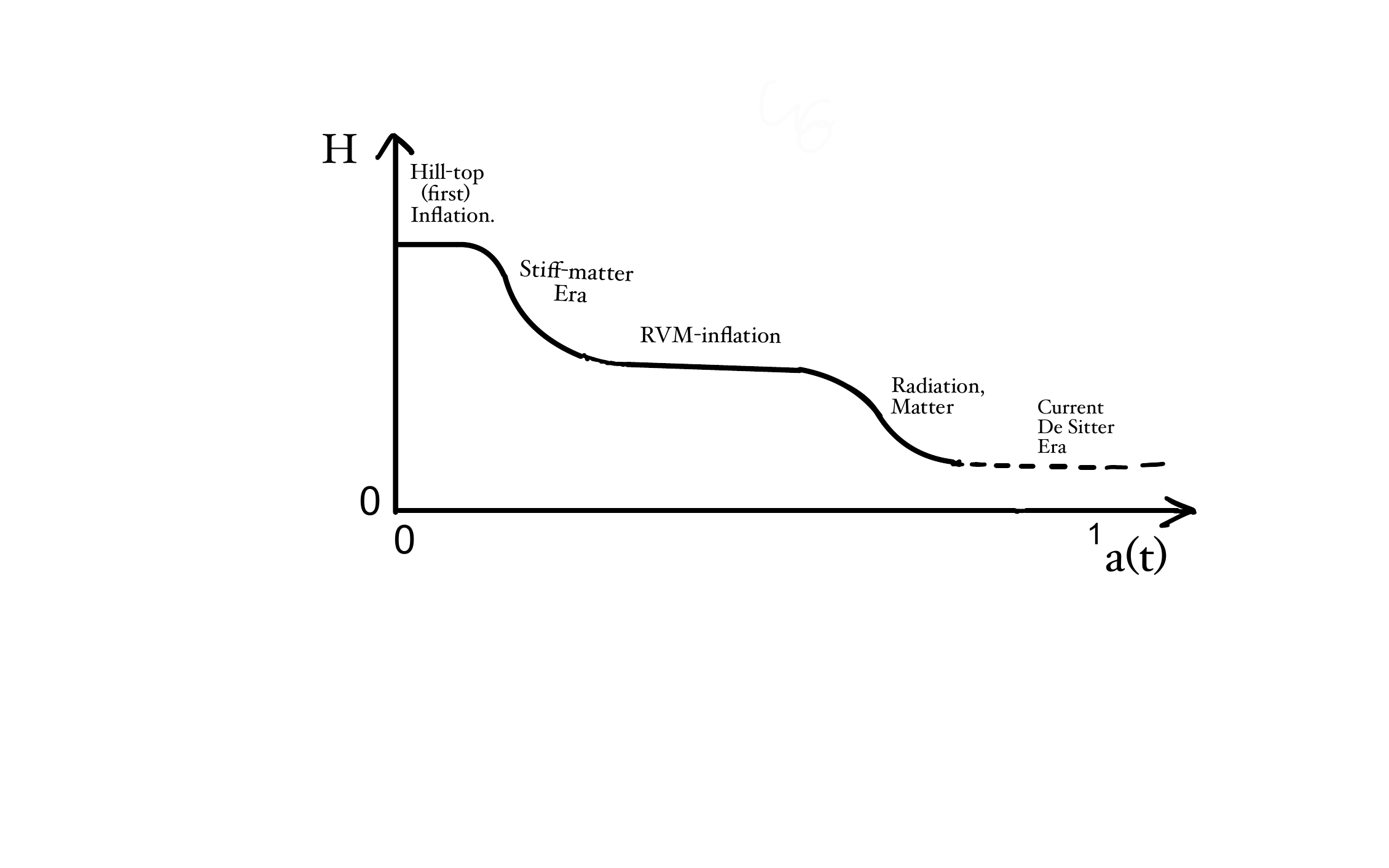}
\vspace{-3.4cm}
\end{center}
\caption{Schematic evolution of the Hubblle parameter ($H$) as a function of the scale factor a(t) in the string-inspired RVM universe~\protect\cite{ms1}, from the Big-Bang (a=0) till the current epoch (a=1, in units of today's scale factor).}
\label{fig:srvmhub}
\end{figure}

Embedding this  simple supergravity model  into our string framework implies~\cite{ms0,ms2} that the exit from the hill-top inflation may be succeeded by dominance of the KR axion matter~\cite{ms2}, which, due to the lack of a potential, constitutes a stiff matter fluid (with equation of  state $w=+1$)~\cite{stiff1,stiff2}. 
This era interpolates between the first-hill-top inflation and the GW-induced RVM inflation (see fig.~\ref{fig:srvmhub}). 
Indeed, the presence of GW as a consequence of domain-wall collapse or collisions, implies the appearance of CP-violating anomalous gravitational Chern Simons terms coupled to the KR axions in the effective action \eqref{sea4}. As the cosmic time elapses, conditions for the condensation of GW can develop, leading to the anomaly condensate \eqref{rrt2}, and the eventual 
RVM inflationary era, driven by the $H^4$ term in \eqref{toten}. 

Before closing this subsection, we would like to discuss some formal properties of the one-loop corrected effective action of supergravity in the de-Sitter background, which we shall make use of in section \ref{sec:modern}. As discussed in 
Ref.~\refcite{houston2} (see also Ref.~\refcite{sugrarvm}), the Euclidean (E) one-loop effective action of the dynamically-broken $N=1$ supergravity model, in a given gauge, evaluated in a de Sitter background, with (one-loop corrected) cosmological constant $1 \gg \Lambda > 0$ reads:
\begin{align}\label{effactionl2}
    \Gamma^{(\rm E)} \simeq S_{\rm cl}-\frac{24\pi^2}{\Lambda^2 }\big(\alpha^F_0+\alpha_0^B
    + \left(\alpha^F_{1}+ \alpha^B_{1}\right)\Lambda 
    +\left(\alpha^F_{2}+ \alpha^B_{2}\right)\Lambda^2+\dots\big)~,
\end{align} 
where the superscripts $B (F)$ denote terms arising from integration of massless (quantum) gravitons (B) (and gravitinos (F)), $f$ is the scale of supergravity (and global suprsymmetry) breaking, 
$\sigma_c < f$,~\cite{houston2} is the value of the gravitino condensate field at the minimum of its effective  potential (see fig.~\ref{fig:sugra}) 
and 
$S_{\rm cl}$ denotes the classical supergravity action with tree-level (bare) cosmological constant $\Lambda_0 < 0$ 
\begin{align}\label{l0tree}
	 \frac{\Lambda_0}{\kappa^2} 	
	 \equiv \sigma_c^2  -f^2 ~,
\end{align}
which is necessarily \emph{negative} 
for reasons of incompatibility of (unbroken) supergravity with de Sitter vacua~\cite{houston,houston2,N1sugra}. The one-loop renormalised cosmological constant $\Lambda$, though, is {\it positive}, due to quantum corrections ({\it cf.} \eqref{LL1}), compatible with dynamically broken supergravity. Taking into account that the Euclidean de Sitter volume is~\cite{tseytlin}  
$24\pi^2/\Lambda^2 \rightarrow \int d^4x \sqrt{\widehat g_{\rm E}} $, 
we may express the effective action \eqref{effactionl2} in a generally covariant form  (using the notation $\widehat \dots$ for quantities evaluated in the background de Sitter space-time):
\begin{align}\label{effactionl3}
\Gamma^{\rm (E)} \simeq&-\frac{1}{2\kappa^2} \int d^4 x \sqrt{\widehat g_{\rm E}}
\left[\left(\widehat R-2\Lambda_1 \right)  +\alpha_1 \, \widehat R+
\alpha_2 \, \widehat R^2\right]~,
\end{align}
where we have replaced $\Lambda$ by the curvature
scalar, 
\begin{align}\label{rl}
\widehat R = 4\Lambda.
\end{align}
The various quantities entering  \eqref{effactionl3} are given by~\cite{houston2,sugrarvm} 
    \begin{align} \label{LL1}
\Lambda_{1}=-\, \kappa^2 \,
\left(-\frac{\Lambda_0}{\kappa^2}+\alpha_0^F+\alpha_0^B\right)~,
    \end{align}
 where \begin{align}\label{a0}
		\alpha_0^{F} &= {\kappa}^4 \, \sigma_c^4 \, \Big(0.100\,  \ln \left( \frac{{\kappa}^2 \, \sigma_c^2}{3 \mu ^2}\right) + 0.126 \Big)~, \nonumber \\
		 				\alpha_0^{B} &= \kappa ^4 \, \left(f^2-\sigma_c^2\right)^2 \left(0.027 - 0.018 \ln \left(\frac{3 \kappa ^2 \left(f^2-\sigma_c^2\right)}{2 \mu ^2}\right)\right)~,
		\end{align}
		and
\begin{align}\label{alpha}
         \alpha_1=\frac{\kappa^2}{2}\left(\alpha^F_1+\alpha^B_1\right)~,\quad
        \alpha_2=\frac{\kappa^2}{8}\left(\alpha^F_2+\alpha^B_2\right)~,
    \end{align}
    with
\begin{eqnarray}\label{aif}
\alpha^F_1&=& 0.067\, \kappa^2 \sigma_c ^2  -0.021\,
\tilde\kappa^2 \sigma_c ^2 \, {\rm ln}
\left(\frac{\Lambda}{\mu^2}\right)
 +  0.073\, \kappa^2 \sigma_c ^2 \, {\rm ln}
\left(\frac{\kappa^2\sigma_c^2}{\mu^2} \right)~,
\nonumber \\
\alpha^F_{2}&=& 0.029 + 0.014\, {\rm ln} \left(\frac{\kappa^2\sigma_c^2}{\mu^2}\right) 
-0.029\, {\rm ln} \left(\frac{\Lambda}{\mu^2}\right)~, \nonumber \\
\alpha^B_1&=& -0.083 \Lambda_0 + 0.018\, \Lambda_0 \, {\rm ln} \left(\frac{\Lambda }{3 \mu ^2}\right)  + 
0.049\, \Lambda_0\,  {\rm ln} \left(-\frac{3 \Lambda_0}{\mu
^2}\right)~, \nonumber \\ \alpha^B_{2} &=& 0.020 +  0.021\, {\rm ln} \left(\frac{\Lambda }{3 \mu ^2}\right) - 
0.014\, {\rm ln} \left(-\frac{6 \Lambda_0}{\mu ^2}\right)~,
\end{eqnarray}
with the replacement \eqref{rl} understood in all the above expressions.

The scale $\mu^2$ is a proper-time ultraviolet cutoff, used to regularise UV divergences~\cite{houston,houston2,tseytlin}. Supergravity breaks dynamically for $\mu^2$ close to Planck scale~\cite{houston}. 
We then set  from now on
\begin{align}\label{muPl}
\mu^2 \sim M^2_{\rm Pl} = \kappa^{-2}.
\end{align}
It can be arranged that the masses of the gravitino condensate and the gravitino become very large compared to the low-energy mass scales. They can be taken to be above the grand-unification scale, even close to Planck scale, for our purposes~\cite{ms1,ms2,ms0}, provided the scale $f$ is chosen appropriately. 
Since the effective action \eqref{effactionl3} is evaluated after one-loop integration of (weak) graviton perturbations about the de Sitter 
background, it is actually a quantum-gravity effective action. 

In the broken supergravity RVM phase, where the hypermassive gravitino and gravitino-condensate fields, with masses of order of the Planck mass, are integrated out from the low-energy effective action, leading to Planck-mass suppressed terms, 
the only leading quantum corrections are those due to massless graviton integration. As can be seen from the expressions \eqref{LL1},\eqref{a0},\eqref{alpha},\eqref{aif}, such corrections involve $\widehat R^{n}\, {\rm ln}(\kappa^2 \widehat R)$, $n=1,2$ terms ({\it cf.} coefficients with superscript $B$).  Passing onto a slowly-varying global de Sitter background (via the coordinate transformations~\cite{lanczos}  \eqref{lanczostr}), we may approximately replace $\widehat R \rightarrow 12 H^2$, where $H$ is the approximately constant inflationary Hubble parameter. 

In the framework of our cosmology, this RVM phase will also involve
the KR axion and gravitational anomaly terms,  
in addition to the aforementioned quantum-gravity corrections of the effective action \eqref{effactionl3} contributing terms of the form $H^{2 n'} {\rm ln}(\kappa^2 H^2)$, $n'=1,2$. From the pertinent expressions \eqref{LL1},\eqref{a0},\eqref{alpha},\eqref{aif}, it becomes clear that, on assuming $\kappa^2 |\Lambda_0| < 1$, for the scale of the bare cosmological constant \eqref{l0tree}, and taking into account \eqref{b0},  
these quantum-gravity-induced corrections are subdominant compared to the 
gravitational-anomaly-condensate-induced $H^4$ term in \eqref{toten}. Thus, our discussion and conclusions on the r\^ole of anomalies in inducing an RVM inflation are not affected~\cite{ms0,ms1}. 

As we speculated though in Refs.~\refcite{ms1,ms0}, quantum-gravity corrections of the form $H^2 {\rm ln}(\kappa^2 H^2)$ could lead to observable deviations from the $\Lambda$CDM in the modern era (which again is characterised by an approximately de Sitter cosmological space-time background, with $H \simeq H_0$). In particular, they could lead to a modified version of RVM, somewhat analogous to the type-II RVM considered in Ref.~\refcite{solaepl} for the 
simultaneous alleviation of the $H_0$ and $\sigma_8$ tensions. We shall briefly discuss these issues in section \ref{sec:modern}.

\section{After RVM inflation: KR-Axion  Dark Matter and Leptogenesis \label{sec:postinfl}}

In this section we review briefly the situation charactrerising the exit from the RVM inflation and 
post inflationary eras of the model. For details we refer the reader to Refs.~\refcite{basilakos,basilakos2,ms1,ms2}. 

At the exit from RVM inflation, gauge fields and chiral fermionic matter, including massive sterile right-handed neutrinos (RHN) of interest to us here, are assumed to be generated from the decay of the RVM vacuum~\cite{basilakos}. Chiral fermions have their own gravitational and chiral anomalies. The former are assumed~\cite{basilakos2,ms2} to cancel the primordial gravitational anomalies due to the Green-Schwarz counterterms. The chiral anomalies, however, remain in the radiation and matter eras~\cite{basilakos}:
\begin{align}\label{anom2}
& \partial_\mu \Big[\sqrt{-g}\, \Big(  \sqrt{\frac{3}{8}} \frac{\alpha^\prime}{\kappa}\, J^{5\mu}  -  \frac{\alpha^\prime}{\kappa}\, \sqrt{\frac{2}{3}}\,
\frac{1}{96} \, {\mathcal K}^\mu  \Big) \Big]     \nonumber \\ 
& = \sqrt{\frac{3}{8}} \, \frac{\alpha^\prime}{\kappa}\, \Big(\frac{\alpha_{\rm EM}}{2\pi}  \, \sqrt{-g}\,  {F}^{\mu\nu}\,  \widetilde{F}_{\mu\nu}
+ \frac{\alpha_s}{8\pi}\, \sqrt{-g} \, G_{\mu\nu}^a \, \widetilde G^{a\mu\nu} \Big)~, \qquad 
\end{align}
where $J^{5\mu} = \sum_{i=\rm chiral~fermions}\overline \psi_i \gamma^5 \gamma^\mu \psi_i$, $F_{\mu\nu}$ denotes the electromagnetic (EM) Maxwell tensor,  $G_{\mu\nu}^a$ is the gluon field strength, with $a=1, \dots 8$ an adjoint SU(3) colour index, $\alpha_{\rm EM}$ the electromagnetic fine structure constant, and $\alpha_s$ the strong-interactions fine structure constant. The gluon terms in \eqref{anom2} may lead, through non-perturbative instanton effects in the post-inflationary QCD era,  to the generation of a potential and a mass $m_b$ term, for the KR axion, which, due to \eqref{stringscale}, 
 is within phenomenologically acceptable ranges~\cite{basilakos2,ms0}, 
 \begin{align}\label{KRmass}
 1.17 \times 10^{-8} \lesssim m_b/(\rm eV) \lesssim 1.17 \times 10^{-5}.
 \end{align}
In this way, the KR axion field can play the r\^ole of a dominant component of dark matter~\cite{basilakos}. In view of the association of the KR axion with torsion, then, one obtains a geometric origin of the dark matter sector of this Universe.

Moreover, the spontaneous-LV KR axion background \eqref{slowrollbint}, which remains undiluted at the exit phase from RVM inflation, can trigger the generation of Leptogenesis in models involving RHN, according to the mechanism discussed in Refs. 
\refcite{sarkar3}, \refcite{sarkar6}. The pertinent lagrangian of the RHN and their interactions with the standard model (SM) sector is:
\begin{align}\label{smelag}
\mathcal{L}= {\mathcal L}_{\rm SM} + i\overline{N}\, \gamma^\mu\, \partial_\mu \, N-\frac{m_N}{2}(\overline{N^{c}}N+\overline{N}N^{c})-\overline{N}\gamma^\mu\, B_\mu \, \gamma^{5}N-\sum_f \, y_{f}\overline{L}_{f}\tilde{\phi}^dN+ {\rm h.c.}
\end{align}
where h.c.  denotes hermitian conjugate, ${\mathcal L}_{\rm SM}$ denotes the SM Lagrangian,
$N$ is the RHN field, of (Majorana) mass $m_N$,  $\tilde \phi$ is the SU(2) adjoint of the Higgs field  $\phi$ ($\tilde{\phi}^d_i \equiv \varepsilon_{ij}\phi_j~, \, i,j=1,2,$ SU(2) indices),
$L_{f}$ is a SM  lepton doublet, with $f$ a generation index, $f=e, \mu, \tau$, $y_f$ is a Yukawa coupling of the ``Higgs portal'' interactions between the SM sector and RHN, which can lead to the generation of SM light neutrino masses, via, {\it e.g.},  the seesaw mechanism. The  quantity $B_\mu$ is associated with the KR axion background \eqref{slowrollbint}
 induced by the anomaly condensates during the RVM inflation:
\begin{align}\label{background}
B_\mu = M_{\rm Pl}^{-1} \, \dot{\overline b}\, \delta_{\mu0}\,.
\end{align}
As discussed in Ref.~\refcite{basilakos2}, during the radiation era, $B_\mu$ is slowly varying with the cosmic temperature $T$ ($B_0 \propto T^3$), so
the Lagrangian (\ref{smelag}) resembles that of a Standard Model Extension (SME) Lagrangian in a LV and CPT Violating (CPTV) (approximately constant) background~\cite{sme,smebounds}. A lepton asymmetry is then generated due to asymmetric 
tree-level decays of the RHN $N$ into SM leptons and anti-leptons in the background \eqref{background} between the channels I and II:~\cite{sarkar3,sarkar6} ${\rm Channel} ~I: \,  N \rightarrow l^{-}h^{+}~, ~ \nu \, h^{0}$,  and 
${\rm Channel ~II}: \, N \rightarrow l^{+}h^{-}~,~  \overline \nu \, h^{0}$,
where $\ell^\pm$ are charged leptons, $h^\pm$ are the charged Higgs fields, which, at the high temperatures, above the electroweak symmetry breaking, at which this leptogenesis takes place~\cite{sarkar3,sarkar6}, 
do not decouple from the physical spectrum, $\nu$ ($\overline \nu$) are the light SM neutrinos (antineutrinos), and $h^0$ is the neutral Higgs field. The lepton asymmetry reads (assuming that the dominant lepton asymmetry is generated by one sterile neitrino, although extension to more RHN species, as required by the seesaw mechanism, is straightforward)~\cite{sarkar6}: 
 \begin{align}\label{lepto}
 \frac{ \Delta L^{TOT}(T=T_D)}{s} \sim  q\,  \dfrac{B_{0}(T_D)\, m_N^2}{T_D^3} \sim \, q\, 3.5 \times 10^{11} \,
 \Big(\dfrac{m_N}{M_{\rm Pl}}\Big)^2, \quad q > 0,
 \end{align}
 where $s \propto T^3$ is the entropy density of the Universe,  $T_D \simeq m_N$ denotes the freeze-out temperature, and $0 < q=\mathcal O(10)$ is a numerical factor expressing theoretical uncertainties in the approximate analytic (Pad\'e) methods used~\cite{sarkar6}. The lepton asymmetry \eqref{lepto}  can then be communicated to the baryon sector via Baryon-minus-Lepton-number ($B-L$) conserving sphaleron processes in the SM sector.

\section{Modern Era Phenomenology:  links of the stringy RVM to a modified type-II RVM and potential data tension alleviation \label{sec:modern}}

In the modern era, the energy density of this stringy Universe also assumes an RVM form, 
\begin{align}\label{modernDE}
\rho_0 = \frac{3}{\kappa^2} \Big(c_0 + \nu_0 \, H_0^2 \Big)~, 
\end{align}
where now $\nu_0 > 0$, due to contributions from cosmic electromagnetic background fields~\cite{basilakos2} (we ignore terms of order $H_0^4$, as they are negligible in the current era). Comparison with the data~\cite{rvmevol2,rvmsola} indicates
$\nu_0 = \mathcal O (10^{-3})$. The present-day cosmological constant $c_0 > 0$ cannot be uniquely determined in our string-inspired cosmology, given that there might be various contributions to it~\cite{ms0}, some associated with details of the extra dimensional geometry of the underlying microscopic string/brane theory. For comparison with data, we therefore should treat $c_0$ as a phenomenological parameter at this stage.

What is important to realise is that, as discussed in subsection \ref{sec:sugra}, integrating out massless quantum graviton fluctuations in the path integral of the current-era gravitational theory, 
which is also characterised by a de Sitter background, generates terms in the one-loop effective action of the form (taking into account \eqref{muPl}):
\begin{equation}\label{1looplagr}
\delta \mathcal L^{\rm 1-loop}_{\rm quant.~grav.} = \sqrt{-\widehat g}\, \Big[ \widetilde \alpha_0 + \widehat R \Big(c_1 + c_2\, {\rm ln}(\frac{1}{12}\kappa^2 \widehat R)\Big) \Big] + \dots 
\end{equation}
From the supergravity example~\cite{houston,houston2,sugrarvm} we have that 
 the constant coefficients $c_i \propto \kappa^2 \mathcal E_0, \, {\rm or} \, c_i \propto \kappa^2 \mathcal E_0 \, {\rm ln}(\kappa^4 |\mathcal E_0|)$, $ i=1,2,$ ({\it cf.} \eqref{aif}) with $\mathcal E_0$ a bare (constant) vacuum energy density scale (the ellipses $\dots $ denote terms of quadratic and higher order in $\widehat R = 12 H^2$, which are subdominant in the current epoch $(H = H_0)$). The structures \eqref{1looplagr} appear generic for weak quantum gravity corrections about de Sitter backgrounds~\cite{tseytlin}, and we may therefore conjecture that they can play a r\^ole in the current era phenomenology. 

 From the graviton equations of the one-loop corrected effective Lagrangian, it can be readily seen that the terms \eqref{1looplagr} will imply corrections to the effective stress-energy tensor in the current era of the form~\cite{ms0}, 
\begin{equation}\label{1loopenden}
\delta \rho_0^{\rm vac}  =  \frac{1}{2}\widetilde \alpha_0 + 3 (c_1-c_2) H_0^2 + 3 c_2 H_0^2 \, {\rm ln}(\kappa^2 H_0^2)~, 
\end{equation}
which should be added to \eqref{modernDE}. Moreover, the supergravity prototype~\cite{houston,houston2,sugrarvm} indicates that the one-loop correction (dark-energy-type) term $\frac{1}{2} \widetilde \alpha_0$ is constant, independent of ${\rm ln} H^2$ terms. This will lead to some crucial differences from the standard type II RVM used in Ref.~\refcite{solaepl} to alleviate the $H_0$ and $\sigma_8$ tensions, which is characterised only by a mild cosmic-time $t$ dependence of an effective gravitational constant, $\kappa^2_{\rm eff} (t) = \kappa^2/\varphi(t)$, with $\varphi(t)$ a phenomenological non-dynamical function. 

Another issue to be clarified is the sign of the bare cosmological constant $\Lambda_0 \propto \mathcal E_0 \, \kappa^2 $
which, as in the supergravity example of subsection \ref{sec:sugra}, enters the computation of the effective action about de Sitter backgrounds in generic quantum gravity (QG) models that could describe the current-era. 
Such a sign depends on the details of the underlying QG theory. In the supergravity example this is negative \eqref{l0tree} (Anti de Sitter type (AdS)), due to fundamental reasons of compatibility with supersymmetry~\cite{houston,houston2}. One may also use such negative bare cosmological constant scales as formal regulators 
in the quantization procedure, 
as, {\it  e.g.}, has been considered in the context of black hole physics~\cite{winst} motivated by the AdS/conformal-field-theory correspondence~\cite{ads}. We stress though that the one-loop corrected cosmological constant is always positive in such constructions, compatible with the current phenomenology (and also with the  broken supergravity scenario~\cite{houston,houston2,sugrarvm}). 
We plan to study such issues, and their modern-era phenomenology, in future works.

Before closing the section, we mention that in string theory there might be a mixing of the KR axion with the other,
string-model dependent, axions that exist as a result of the compactification procedure~\cite{svrcek}. In Ref.~\refcite{basilakos2}, we have argued that, depending on details of this mixing, one might obtain 
KR axion backgrounds that may be characterised by an approximately constant $\dot b$ even in the current era of the 
string Universe. Such backgrounds can therefore be parametrised by
\begin{align}\label{modernbdot}
\dot b_0 \sim \sqrt{2\, \epsilon^\prime} \, H_0\, M_{\rm Pl}.
\end{align}
Phenomenological considerations, associated with the r\^ole of the KR axion background as dark matter, imply then  
that $\epsilon^\prime \sim 10^{-2}$, in order to reproduce a dark matter energy density due to the KR axion in the
phenomenologically right ballpark~\cite{pdg,Planck}. This, in turn, leads to 
a curious coincidence  in the order of magnitude of the slow-roll parameters entering the expressions for $\dot b$ in the RVM-inflationary \eqref{slowrollbint} ($\epsilon$) and modern \eqref{modernbdot} eras ($\epsilon^\prime$), 
\begin{align}\label{coinc}
\epsilon^\prime \sim \epsilon \simeq 10^{-2}~, 
\end{align}
which needs to be understood further, in the context of the underlying microscopic string theory models.  We do note though~\cite{basilakos,ms0} that
the background \eqref{modernbdot}, under the assumption \eqref{coinc}, is compatible with the stringent current experimental bounds of LV and CPTV discussed in ref.~\refcite{smebounds}.

\section{Conclusions and Outlook \label{sec:concl}} 

In this talk, I reviewed a string-inspired model of a running vacuum (RVM) cosmology, which seems consistent with the current phenomenology, but also provides a geometrical origin of both, RVM inflation and dark matter. Crucial r\^oles are played by the gravitational anomalies that characterise the model at early eras, 
and the KR axion fields from the fundamental massless multiplet of the underlying string theory. The KR axion field is associated with a totally antisymmetric torsion in the underlying string theory. The RVM inflation arises from  the non-linear $H^4$ term in the vacuum energy density \eqref{toten}, which is exclusively due to primordial GW that induce non trivial condensates of the gravitational Chern-Simons terms that are present in the very early stages of this string-inspired cosmology. Such terms do not arise in the context of anomaly-free quantum field theories, or Starobinsky inflation, and, as such, are rather exclusive to our string-inspired RVM. The anomaly condensates also imply LV  KR axion backgrounds that induce baryogenesis through leptogenesis. In this sense, we may 
dare state that our very existence, that  is the dominance of matter over antimatter in the Cosmos, is due to cosmic
anomalies. {\it We are anomalously made of star stuff}, to 
paraphrase the famous quote by Carl Sagan~\cite{bas0,basilakos2}. 

There are several open issues that we need to understand, and explore further, such as: (i) potential  hints in the cosmic-microwave-background data~\cite{Planck} about the LV KR axion backgrounds, (ii) prospects of getting some phenomenological
indications in early-Universe data about the negative, anomaly-induced, coefficient of the $H^2$ term in the RVM-like cosmic energy density \eqref{toten} during the RVM inflation. This, however, is a difficult task, in view of the phase transition occurring in our model at the exit from the RVM inflation,
(iii) the rich phenomenology of both, the string-model-independent KR axion and the other, string-model dependent, axions, that exist in string theory~\cite{svrcek,arvanitaki},
(iv) the precise r\^ole of quantum-gravity-induced $H^2\, {\rm ln}(\kappa^2 H^2)$ corrections in the vacuum energy density of the current era, comparing the model with the type-II RVM~\cite{solaepl}, in an attempt to discuss potential alleviation of the observed tensions in the cosmological data, and (v) last but not least: in view of the association of the KR axion with (totally antisymmetric) torsion, the phenomenological comparison of the stringy RVM with 
other cosmologies with torsion~\cite{tors1,tors2} (in which the torsion has more components than the totally antisymmetric one characterising our model) or other teleparallel-gravity models~\cite{tele} (where torsion mimics gravity, in contrast to our case, which contains also graviton fields). We hope to be able to report on these important issues in the near future.

\vspace{-0.1cm} 
\section*{Acknowledgements}

NEM thanks A. G\'omez-Valent and J. Sol\`a Peracaula for the invitation to speak in the {\it CM3} parallel session: 
``{\it Status of the $H_0$ and $\sigma_8$ tensions: theoretical models and model-independent constraints}'' 
of the MG16 Marcel Grossmann virtual Conference, July 5-10 2021. He also acknowledges participation in the COST Association Action CA18108 ``{\it Quantum Gravity Phenomenology in the Multimessenger Approach (QG-MM)}''. This work is funded in part by the UK Science and Technology Facilities  research Council (STFC) under the research grant ST/T000759/1.

\end{document}